\documentclass[prl,aps,twocolumn,superscriptaddress]{revtex4}

\usepackage[dvips]{graphicx}
\usepackage{amssymb,amsfonts,amsmath}
\usepackage{color}
\usepackage{ulem}

\newcommand{\rv}{{\mathbf r}}

\newcommand{\Jv}{{\bf J}}

\newcommand{\Fv}{{\bf F}}
\newcommand{\fv}{{\bf f}}
\newcommand{\Xv}{{\bf X}}

\newcommand{\vel}{{\bf v}}

\newcommand{\rnt}{{\bf r}^N\!\!,t}

\begin{document}

\title{Velocity gradient power functional for Brownian dynamics}

\author{Daniel de las Heras}
\affiliation{Theoretische Physik II, Physikalisches Institut, 
  Universit{\"a}t Bayreuth, D-95440 Bayreuth, Germany}

\author{Matthias Schmidt}
\affiliation{Theoretische Physik II, Physikalisches Institut, 
  Universit{\"a}t Bayreuth, D-95440 Bayreuth, Germany}

\date{3 October 2017}

\begin{abstract}
We present an explicit and simple approximation for the superadiabatic
excess (over ideal gas) free power functional, admitting the study of
the nonequilibrium dynamics of overdamped Brownian many-body
systems. The functional depends on the local velocity gradient and is
systematically obtained from treating the microscopic stress
distribution as a conjugate field. The resulting superadiabatic forces
are beyond dynamical density functional theory and are of viscous
nature. Their high accuracy is demonstrated by comparison to
simulation results.
\end{abstract}

\maketitle

The response of complex systems to external stresses is important,
both from an applied point of view of control of flow properties, but
also from a fundamental interest in understanding the collective
nonequilibrium behaviour of many-body systems
\cite{brader10review}. In particular colloidal dispersions, when
exposed to shear flow, display a wealth of striking nonequilibrium
phenomena, ranging from the nonlinear rheological behaviour of fluids
\cite{brader10review} and glasses \cite{fuchs2017overview} to shear
banding phase transitions \cite{dhont1999,dhont2003,dhont2014}. Much
theoretical work has been carried out on the basis of the Smoluchowski
many-body equation for overdamped Brownian systems
\cite{brader10review}.  On its basis dynamical density functional
theory (DDFT)
\cite{evans79,marinibettolomarconi99,archer04ddft,evans2016specialIssue}
has been used in order to study rheological properties of model
fluids. Brader, Kr\"uger and their coworkers
\cite{braderkrueger11epl,braderkrueger11molphys,reinhardt2013,aerov2014,aerov2015,scacchi2016}
have supplemented the DDFT by further physically motivated
contributions, such as e.g.\ kinetic (flow kernel) considerations, in
order to address a range of specific rheological problems.  Their
approach relies on considering two-point distribution functions, which
they incorporate into DDFT.

The power functional theory (PFT) \cite{power} is a formally exact and
general dynamical approach which rather operates on the level of the
time-dependent one-body density, $\rho(\rv,t)$, and the current
distribution, $\Jv(\rv,t)$. A minimization principle determines the
current at position $\rv$ and time $t$, and hence the time evolution
of the system. The many-body problem is encapsulated in the excess
(over ideal gas) superadiabatic (over free energy) contribution to the
free power functional, $P_t^{\rm exc}[\rho,\Jv]$, which is in general
a spatially and temporally nonlocal functional of both density and
current. The resulting (superadiabatic) forces are obtained from
functional differentiation of $P_t^{\rm exc}$ with respect to
$\Jv(\rv,t)$, upon holding $\rho(\rv,t)$ fixed. The superadiabatic
forces act in addition to the adiabatic forces; the latter are
generated from the (equilibrium) free energy functional. On the basis
of PFT, nonequilibrium Ornstein-Zernike relations
\cite{brader13noz1,brader13noz2} were systematically derived.
However, the central object of the theory, $P_t^{\rm exc}$, remains to
a large extent unknown at present, which hampers the application of
PFT to concrete problems.

In this paper, we construct an explicit approximation for $P_t^{\rm
  exc}$, based on a re-formulation of PFT using the local velocity
gradient field and the microscopic stress tensor as its conjugate
field. This enables us to (i) connect PFT to rheology, and (ii)
systematically construct an approximate superadiabatic excess
functional. As we demonstrate, in rheological problems the
superadiabatic forces describe viscous effects. These can be large and
can even be the sole effects present, i.e.\ in cases where the
adiabatic effects vanish, such as in bulk steady shear flow. Hence
rather than merely correcting DDFT, our current approach offers the
study of entirely distinct areas of physics. To test the
validity of our approach, we study the time evolution of a system of
hard particles and find excellent agreement between theory and
Brownian dynamics simulation results.

The starting point of PFT is a generator on the many-body level
\cite{power}, defined as
\begin{align}
  {\cal R}_t =  \int d\rv^N \Psi(\rnt)\sum_i
  \Big(\frac{\gamma}{2}\tilde\vel_i^2
  -\tilde\vel_i\cdot\Fv_i
   + \dot V_{{\rm ext},i}\Big),
  \label{EQRtManyBody}
\end{align}
where $\gamma$ is the friction constant, $\Psi(\rv^N,t)$ is the
time-dependent many-body probability distribution in configuration
space of $N$ particles, spanned by $\rv^N\equiv\{\rv_1\ldots\rv_N\}$,
where $\rv_i$ is the position coordinate of particle $i=1\ldots N$,
$\tilde\vel_i(\rv^N,t)$ is the trial velocity function of
particle~$i$, $\Fv_i(\rnt)$ is the total force acting on particle $i$,
and $\dot V_{{\rm ext},i}\equiv\partial V_{\rm ext}(\rv_i,t)/\partial
t$ is the partial time derivative of the external one-body potential
$V_{\rm ext}$. The ``real'' velocity $\vel_i(\rv^N,t)$ of particle $i$
arises, in the over-damped limit considered here, as
\begin{align}
  \gamma \vel_i(\rnt) &= {\bf F}_i(\rnt),
  \label{EQofMotion}
\end{align}
where
\begin{align}
  {\bf F}_i(\rnt) &= -k_{\rm B}T \nabla_i \ln \Psi(\rnt)
  -\nabla_i u(\rv^N)
\notag\\
  &\quad - \nabla_i V_{\rm ext}(\rv_i,t) + {\bf X}(\rv_i,t)
  +\gamma \vel_{\rm sol}(\rv_i,t),
  \label{EQforceDefinition}
\end{align}
with $k_{\rm B}$ being the Boltzmann constant, $T$ is absolute
temperature, $\nabla_i$ indicates the derivative with respect to
$\rv_i$, $u(\rv^N)$ is the inter-particle interaction potential,
$\Xv(\rv,t)$ is a non-conservative external force field, and
$\vel_{\rm sol}(\rv,t)$ is the imposed velocity field of the
(implicit) solvent; here $\rv$ is the space coordinate.  The many-body
(free power) functional \eqref{EQRtManyBody} is constructed in such a
way that minimization with respect to all $\tilde\vel_i(\rnt)$, which
implies that $\partial {\cal R}_t/\partial \tilde \vel_i(\rnt)=0$ at
the minimum, sets each trial velocity equal to the corresponding real
velocity, $\tilde\vel_i(\rnt)=\vel_i(\rnt)$. This process is carried
out at each point in time, and the resulting dynamics for $\Psi(\rnt)$
is equal to that given by the Smoluchowski equation \cite{power}.  The
many-body functional \eqref{EQRtManyBody} is significant as it acts as
a generator of averages of interest, with one (primary) example being
$\delta {\cal R}_t/\delta {\bf X}(\rv,t)=-\Jv(\rv,t)$, evaluated at
the minimum, where the one-body current distribution is the
microscopic average
\begin{align}
  \Jv(\rv,t) &= \int d\rv^N \Psi(\rnt) \sum_i 
  \delta(\rv-\rv_i) \vel_i(\rv^N,t),
  \label{EQcurrentDefinition}
\end{align}
with $\delta(\cdot)$ being the (three-dimensional) Dirac distribution.

Here we start by considering the functional derivative of ${\cal R}_t$
with respect to the velocity gradient of the solvent, and obtain
\begin{align}
  \frac{\delta {\cal R}_t}{\delta \nabla \vel_{\rm sol}(\rv,t)}
  &= \boldsymbol\sigma(\rv,t), \label{EQRtManyBodyYieldsSigma}
\end{align}
where the local and time-resolved stress distribution
$\boldsymbol\sigma(\rv,t)$ is a one-body second-rank tensor field.
Any microscopic definition of $\boldsymbol\sigma(\rv,t)$ is
necessarily non-unique \cite{schofield82}, as can be gleaned from the
fact that the (driving) force density is obtained via the divergence,
\begin{align}
 \nabla \cdot\boldsymbol\sigma(\rv,t) &=  \gamma \Jv(\rv,t),
  \label{EQforceDensityFromSigma}
\end{align}
where $\Jv(\rv,t)$ is the average
\eqref{EQcurrentDefinition}. Clearly, \eqref{EQforceDensityFromSigma}
is invariant under adding a divergenceless tensor field to
$\boldsymbol\sigma(\rv,t)$. In practice, carrying out the derivative
\eqref{EQRtManyBodyYieldsSigma} of \eqref{EQRtManyBody} requires to
specify an inversion operation to $\nabla$. For simplicity, we choose
this to be the convolution with a radial, inverse square distance
vector field,
\begin{align}
  \nabla^{-1} f(\rv) &= \int d\rv'
  \frac{\rv-\rv'}{4\pi|\rv-\rv'|^3}f(\rv'),
  \label{EQinverseNablaDefinition}
\end{align}
where $f(\cdot)$ is a test function.  $\nabla\cdot\nabla^{-1}
f(\rv)=f(\rv)$ is indeed the identity, as can easily be checked upon
exploiting the identity $\delta(\rv)=\nabla\cdot[\rv/(4\pi|\rv|^3)]$.

The specific form of $\boldsymbol\sigma(\rv,t)$ then emerges as a
microscopic average from \eqref{EQRtManyBodyYieldsSigma} upon spatial
integration by parts,
\begin{align}
  \boldsymbol\sigma(\rv,t) &=
  \int d\rv^N \Psi(\rnt) \sum_i 
  \frac{(\rv-\rv_i){\bf F}_i(\rv^N,t)}{4\pi |\rv-\rv_i|^3},
  \label{EQstressTensorMicroscopic}
\end{align}
where the vector product on the right hand side is a dyadic.  For the
special case of pairwise interparticle forces, the form
\eqref{EQstressTensorMicroscopic} was suggested by Wajnryb et
al.\ \cite{wajnryb95}, but apparently not used further. A common
alternative is that given by Irving and Kirkwood
\cite{irvingKirkwood}; however, its extension to higher than two-body
forces becomes increasingly
cumbersome. Eq.~\eqref{EQstressTensorMicroscopic} does not suffer from
this deficiency.

As a consequence of the structure of
\eqref{EQstressTensorMicroscopic}, the force density relationship
\eqref{EQforceDensityFromSigma} follows upon observing the factor
$\gamma$ from \eqref{EQofMotion}.  The stress tensor distribution
\eqref{EQstressTensorMicroscopic} carries further significance, as it
allows us to define an integrated stress $\boldsymbol\Sigma(t)$ via
spatial integration of the stress distribution
$\boldsymbol\sigma(\rv,t)$ over~${\mathbb R}^3$,
\begin{align}
  \boldsymbol\Sigma(t) & = 
  \int d\rv \boldsymbol\sigma(\rv,t) \label{EQstressDistributionIntegral}
  \\ &=
  -\frac{1}{3}\int d\rv^N \Psi(\rnt)\sum_i \rv_i {\bf F}_i(\rnt),
  \label{EQClausiusVirial}
\end{align}
where the form \eqref{EQClausiusVirial} follows from inserting
\eqref{EQstressTensorMicroscopic} into
\eqref{EQstressDistributionIntegral} and carrying out the $\rv$
integral. The negative trace of the stress tensor, $-{\rm
  Tr}\;\boldsymbol\Sigma(t)$, is the (averaged) Clausius virial
\cite{Hansen06}. Eqs.~\eqref{EQforceDensityFromSigma} and
\eqref{EQClausiusVirial} attest to the fact that
\eqref{EQstressTensorMicroscopic} is a meaningful definition of a
general local and time-resolved stress distribution. In the following
we use \eqref{EQstressTensorMicroscopic} in order to formulate power
functional theory on the tensorial level of the microscopic stress
distribution and the velocity gradient.

PFT elevates the variational principle on the
one-body level, via constructing, from \eqref{EQRtManyBody}, a
one-body ``free power'' functional $R_t$ that depends on the one-body
density distribution $\rho(\rv,t)$, and on $\Jv(\rv,t)$, and which can
be split according to
\begin{align}
  R_t &= P_t^{\rm id} + P_t^{\rm exc} +\dot F - X_t,
  \label{EQRtSplitting}
\end{align}
where $P_t^{\rm id}$ is the ideal dissipation functional,
\begin{align}
  P_t^{\rm id}
  &=\int d\rv  \frac{\gamma \Jv(\rv,t)^2}{2\rho(\rv,t)},
\end{align}
and $P_t^{\rm exc}$ is the excess (over ideal) contribution, which
arises from the presence of internal interactions, $\dot F =\int d\rv
\Jv(\rv,t)\cdot\nabla \delta F/\delta \rho(\rv,t)$ \cite{power} is the
total time derivative of the (equilibrium) intrinsic Helmholtz free
energy density functional $F[\rho]$, and $X_t$ is the external power,
given by the simple space- and time-local expression
\begin{align}
  X_t  &= \int d\rv (\Jv(\rv,t)\cdot {\bf f}_{\rm ext}(\rv,t)
  -\rho(\rv,t) \dot V_{\rm ext}(\rv,t)),
  \label{EQ_XtViaCurrent}
\end{align}
where the total external force field is ${\bf f}_{\rm ext}(\rv,t)=
-\nabla V_{\rm ext}(\rv,t)+{\bf X}(\rv,t)+\gamma \vel_{\rm
  sol}(\rv,t)$.  Here $\rho(\rv,t)= \int
d\rv^N\Psi(\rv^N,t)\sum_i\delta(\rv-\rv_i)$ is the microscopic
one-body density distribution.

The variational principle \cite{power} states that $R_t $ is minimized
by the true current at time $t$, at fixed density $\rho(\rv,t)$, which
implies that
\begin{align}
  \left.\frac{\delta R_t}{\delta \Jv(\rv,t)}\right|_{\rho}   &= 0.
  \label{EQvariationalPrincipleCurrent}
\end{align}
The density distribution can then be updated according to the
continuity equation, $\partial \rho(\rv,t)/\partial t =
-\nabla\cdot\Jv(\rv,t)$. Inserting the decomposition
\eqref{EQRtSplitting} into \eqref{EQvariationalPrincipleCurrent}
yields the equation of motion \cite{power}
\begin{align}
  \gamma \vel(\rv,t) &=-k_BT\nabla\ln\rho
  -\nabla\frac{\delta F_{\rm exc}}{\delta\rho(\rv,t)}
  \notag\\&\qquad
  -\left.\frac{\delta P_t^{\rm exc}}{\delta \Jv(\rv,t)}\right|_\rho
  + {\bf f}_{\rm ext}(\rv,t),
  \label{EQofMotionPFT}
\end{align}
where the (negative) friction force (left hand side) is balanced by
the sum of ideal diffusive, excess adiabatic and superadiabatic, and
external forces (right hand side); here the velocity field is defined
as the ratio
\begin{align}
  \vel(\rv,t) &= \Jv(\rv,t)/\rho(\rv,t).
  \label{EQvelocityDefinition}
\end{align}
The excess adiabatic force is $\fv_{\rm adx}(\rv,t)=-\nabla \delta
F_{\rm exc}[\rho]/\delta\rho(\rv,t)$, where the excess (above ideal)
free energy functional $F_{\rm exc}$ is defined via $F[\rho]=F_{\rm
  exc}[\rho]+k_BT\int d\rv \rho(\ln(\rho\Lambda^3)-1)$, where
$\Lambda$ is the (irrelevant) de Broglie wavelength.

Although this (original) formulation of PFT \eqref{EQofMotionPFT}
\cite{power} permits to obtain the full time evolution of the density
and current fields of the system, the stresses that act do not
appear. To provide access, we perform a change of variables,
from the current $\Jv(\rv,t)$ to the gradient of the velocity field,
$\nabla\vel(\rv,t)$. Using \eqref{EQinverseNablaDefinition} and
spatial integration by parts we can rewrite the external power
\eqref{EQ_XtViaCurrent} as
\begin{align}
  X_t &= \label{EQXtStressForm}
  -\int d\rv (\boldsymbol\sigma_{\rm ext}(\rv,t):
  \nabla\vel(\rv,t) 
  + \dot V_{\rm ext}(\rv,t)\rho(\rv,t)),
\end{align}
where the colon indicates a double tensor contraction, and the
external stress is defined as
\begin{align}
  \boldsymbol\sigma_{\rm ext}(\rv,t) &=
  \nabla^{-1}\left[\rho(\rv,t){\bf f}_{\rm ext}(\rv,t)\right].
\end{align}

Due to the structure of \eqref{EQRtSplitting} and
\eqref{EQXtStressForm}, we can generate the velocity gradient tensor
field via functional differentiation,
\begin{align}
  \left.\frac{\delta R_t }
       {\delta \boldsymbol\sigma_{\rm ext}(\rv,t)}\right. &=
       \nabla\vel(\rv,t).
\end{align}
Using the splitting \eqref{EQRtSplitting} further, we also perform
integration by parts to express the ideal and adiabatic
contributions, respectively, as
\begin{align}
  P_t^{\rm id}
  &= -\frac{1}{2}\int d\rv 
  \boldsymbol\sigma(\rv,t):\nabla\vel(\rv,t),
  \label{EQPtidStrainRate}\\
  \dot F
  &= \int d\rv \boldsymbol\sigma^{\rm ad}(\rv,t):
  \nabla \vel(\rv,t),
  \label{EQFdotStrainRate}
\end{align}
where the total stress $\boldsymbol\sigma(\rv,t)$ is a functional of
$\nabla\vel(\rv,t)$ and $\rho(\rv,t)$ via
\eqref{EQforceDensityFromSigma} and \eqref{EQvelocityDefinition}, and
the adiabatic stress $\boldsymbol\sigma^{\rm ad}(\rv,t)$ is given by
\begin{align}
  \boldsymbol\sigma^{\rm ad}(\rv,t) &= 
  -\nabla^{-1}\rho(\rv,t)
  \nabla \frac{\delta F}{\delta \rho(\rv,t)}.
  \label{EQsigmaAdiabatic}
 \end{align}
We can now reformulate the variational principle
\eqref{EQvariationalPrincipleCurrent} as
\begin{align}
  \left.\nabla\cdot\frac{\delta R_t}{\delta \nabla \vel(\rv,t)}
  \right|_\rho &= 0,
  \label{EQvariationalPrincipleStrainRate}
\end{align}
where the density $\rho(\rv,t)$ is kept fixed under the variation.  An
equivalent form is
\begin{align}
  \left.\frac{\delta R_t }{\delta \nabla\vel(\rv,t)}
  \right|_\rho  &= 
  \boldsymbol\sigma_{\rm stat}(\rv,t),
  \label{EQvariationalPrincipleStrainRateWithStatic}
\end{align}
where $\boldsymbol\sigma_{\rm stat}(\rv,t)$ is a ``static'' stress
that generates vanishing force density,
$\nabla\cdot\boldsymbol\sigma_{\rm stat}(\rv,t)=0$.

We next exploit the decomposition \eqref{EQRtSplitting}, and first
consider the velocity gradient form of the ideal dissipation
functional \eqref{EQPtidStrainRate}. Carrying out the functional
derivative (at constant density $\rho(\rv,t)$) yields
\begin{align}
  \left.\frac{\delta P_t^{\rm id}}{\delta \nabla\vel(\rv,t)}\right|_\rho &=
  -\boldsymbol\sigma(\rv,t),
\end{align}
where the factor of $1/2$ from \eqref{EQPtidStrainRate} cancels with
the two possibilities to carry out the integration by parts
(i.e.\ $\boldsymbol\sigma(\rv,t)$ is not kept constant during the
variation).

As the functional derivative of \eqref{EQXtStressForm} 
and of \eqref{EQFdotStrainRate}
is
straightforward, we are now in a position to rewrite
\eqref{EQvariationalPrincipleStrainRateWithStatic} as
\begin{align}
  \boldsymbol\sigma(\rv,t) &=
  \boldsymbol\sigma^{\rm ad}(\rv,t) + 
  \boldsymbol\sigma^{\rm sup}(\rv,t) +
  \boldsymbol\sigma^{\rm ext}(\rv,t) +
  \boldsymbol\sigma^{\rm stat}(\rv,t),
  \label{EQstressBalance}
\end{align}
where the superadiabatic stress tensor $\boldsymbol\sigma^{\rm
  sup}(\rv,t)$ is obtained from the superadiabatic excess functional via
\begin{align}
  \boldsymbol\sigma^{\rm sup}(\rv,t) &\equiv
  \left. 
  \frac{\delta P_t^{\rm exc} }{\delta\nabla\vel(\rv,t)}\right|_\rho
  \\&=
  \left. -\nabla^{-1} \left(\rho(\rv,t) 
  \frac{\delta P_t^{\rm exc} }{\delta \Jv(\rv,t)}\right|_\rho\right).
  \label{EQsigmaSuperAdiabatic}
\end{align}
As a result of the variable transformation between $\Jv$, $\vel$, and
$\nabla\vel$, at fixed density, the excess free power functional can
be alternatively and equivalently expressed as $P_t^{\rm
  exc}[\rho,\Jv]$, $P_t^{\rm exc}[\rho,\vel]$, or $P_t^{\rm
  exc}[\rho,\nabla\vel]$.

The theory laid out so far is an exact reformulation of the many-body
problem in nonequilibrium. Its complexity is entirely contained in the
functional form of $P_t^{\rm exc}$.  It
requires approximations to make further progress. To lowest order in
$\nabla\vel$, we assume a bi-linear form, which is nonlocal in space
and time:
\begin{align}
  P_t^{\rm exc}  &= k_{\rm B}T
  \int d\rv \int d\rv' \int_0^tdt'  \rho(\rv,t) 
  \nabla\vel(\rv,t) \notag\\
  &\qquad :{\sf M}(\rv-\rv',t-t'):\nabla\vel(\rv',t')\rho(\rv',t'),
  \label{EQPtexc_general}
\end{align}
where ${\sf M}(\rv,t)$ is a fourth-rank tensor that carries no
physical units and depends in general functionally on the density
distribution; the state of the system is assumed to be known at the
initial time $t=0$.

On long time scales and for small inhomogeneities we may further
approximate, and use a Markovian and spatially local
approximation. Due to rotational symmetry we obtain the simple form
\begin{align}
  P_t^{\rm exc} &=
  \frac{1}{2}\int d\rv \rho [
    n_{\rm rot}(\nabla\times\vel)^2
    +n_{\rm div}(\nabla\cdot\vel)^2],
  \label{EQPtexc_local}
\end{align}
where $n_{\rm rot}$ and $n_{\rm div}$ are parameters with units of
${\rm energy}\times{\rm time}$.  Hence the dynamical shear and volume
viscosity are given, respectively, by
\begin{align}
  \eta &= \rho n_{\rm rot},\quad
  \zeta = \rho n_{\rm div},
\end{align}
with units of $\rm Pas=Ns/m^2=Js/m^3$. When starting from
\eqref{EQPtexc_general} the viscosities can then be obtained as
moments of the memory kernel {\sf M}.  The full (fourth-rank)
viscosity tensor ${\boldsymbol\eta}(\rv,\rv',t,t')$ is obtained as the
functional derivative
\begin{align}
  \boldsymbol\eta  &= 
  \frac{\delta\boldsymbol\sigma^{\rm sup}(\rv,t)}
       {\delta\nabla\vel(\rv',t')}
       = \left.\frac{\delta^2 P_t^{\rm exc}}
       {\delta \nabla\vel(\rv',t')\delta\nabla\vel(\rv,t)}\right|_\rho.
\end{align}

Assuming constant viscosities and density, the superadiabatic force
field that follows from \eqref{EQPtexc_local} has the  familiar Stokes
form of hydrodynamics \cite{Hansen06}:
\begin{align}
  {\bf f}_{\rm sup}(\rv,t)&\equiv
  -\rho^{-1} \frac{\delta P_t^{\rm exc}}{\delta \vel(\rv,t)} 
  \label{EQfsupAsDerivatice}
  \\&=
  -\eta(\nabla^2\vel-\nabla\nabla\cdot\vel)
  +\zeta \nabla\nabla\cdot\vel.
\end{align}
In the more general case, without the above restrictions,
\eqref{EQfsupAsDerivatice} yields
\begin{align}
  \fv_{\rm sup}(\rv,t) &=
  \nabla \rho n_{\rm rot} \cdot\nabla\vel 
  -\nabla \rho n_{\rm rot} \nabla\cdot\vel
  + \nabla \rho n_{\rm div} \cdot \nabla \vel.
\end{align}

As a proof of concept we apply the power functional approach developed
here to a one-dimensional (1D) system of hard particles, and compare
the results to Brownian dynamics (BD) simulations. A 1D system of hard
particles is an ideal test case since the exact equilibrium density
functional is known \cite{percus}.  Hence, differences between the
time evolution predicted by PFT and that obtained with BD simulations
are primarily due to the use of an approximate PFT.  As our system
contains a reduced number of particles, the use of different
statistical ensembles (grand canonical for the derivative of the free
energy in PFT and canonical in BD) might, in principle, be an
additional source of discrepancy between theory and simulations.  To
minimize this effect, we have selected cases for which the equilibrium
density profiles obtained with DFT and BD are very similar.  In other
cases it would be necessary to first obtain the canonical data from
grand canonical density functional theory
~\cite{canonical1,canonical2}.

We study the time evolution of a system of $N$ hard particles of size
$L$ in a box of length $H$ with periodic boundary conditions. The
system is initially in equilibrium in an external potential given by $
V_{\rm ext}(x)=V_0\sin(2\pi xN/H)$, with $x$ the spatial coordinate.
At $t=0$ we switch off the external potential and study the time
evolution both with BD and PFT. Here we model the superadiabatic
excess functional \eqref{EQPtexc_local} by
\begin{equation}
  P_t^{\rm exc}
  =k_{\rm B}T\frac{K(\bar\rho,t)}{2}\int_0^H dx\rho(x,t)
  \left[\partial_x v(x,t)\right]^2,
  \label{pexc1D}
\end{equation}
where the velocity profile is defined via \eqref{EQvelocityDefinition}
and $K$ is a global prefactor (related to the kernel ${\sf M}$) that
depends on the average density $\bar\rho$ and the time $t$ and takes
into account the memory effects.  The superadiabatic force density
$I_{\rm sup}(x,t)$, which is neglected in DDFT, is given by the
functional derivative \eqref{EQfsupAsDerivatice} of $P_t^{\rm exc}$,
multiplied by the one-body density, i.e.\ $I_{\rm sup}=\rho f_{\rm
  sup}$.

We apply the numerical method of Ref.~\cite{andrea} to
measure $I_{\rm sup}(x,t)$ using BD simulations, and compare to the
theoretical results.  As we will see below, memory plays an important
role during the time evolution of the system. We include memory
effects in the time-dependent prefactor $K(\bar\rho,t)$ of $P_t^{\rm
  exc}$, cf Eq.~\eqref{pexc1D}. The explicit dependence of $K$ with
time will be the focus of a future study. Here we are only interested
in the functional form of $P_t^{\rm exc}$ with the velocity
profile. Hence, to compare theory and simulations we 
(i) obtain $I_{\rm sup}$ and the density profile
$\rho(x,t)$ at a given time $t$ using BD simulations, and (ii) use
$\rho(x,t)$ as input of our PFT and find the value of $K$ that best
reproduces the simulation results. In other words, we fit the
amplitude of the superadiabatic force, but nothing else.

\begin{figure}
\includegraphics[width=0.95\columnwidth]{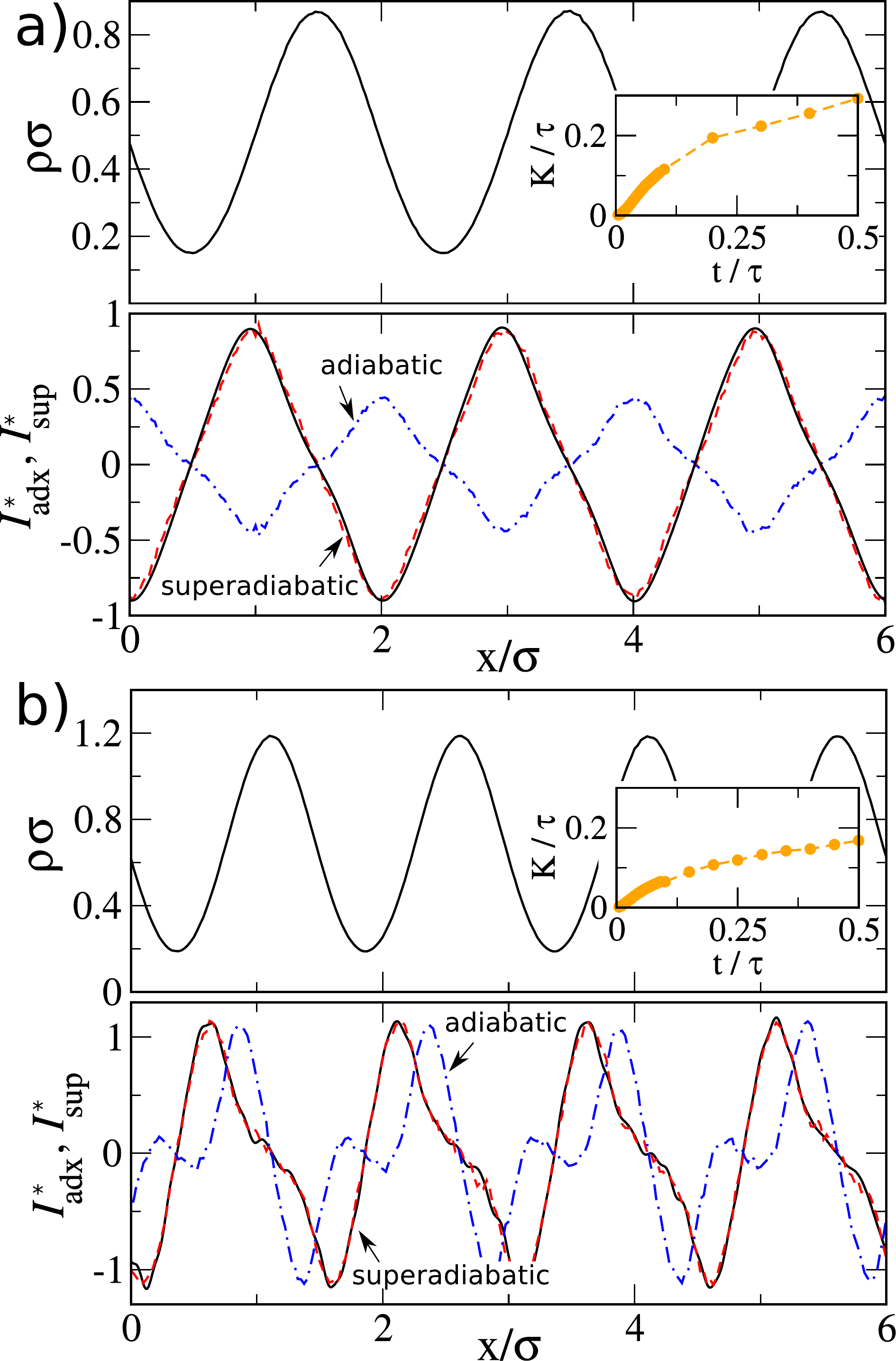}
\caption{(Color online) a) Density profile $\rho(x,t)$ as a function
  of $x$ (top panel) in a periodic system with $N=15$ and size $H=30$
  obtained with Brownian dynamics simulations (only a small portion of
  the box is showed). The bottom panel of a) shows the scaled excess
  (over ideal gas) adiabatic force density
  $I_{\rm adx}^*=\rho f_{\rm adx} L^2/(k_BT)$
  (blue dash-dotted line) as a function of $x$.  The scaled
  superadiabatic force density $I_{\rm sup}^*=I_{\rm sup}L^2/(k_BT)$
  is also shown according to Brownian dynamics simulation (red dashed
  line) and the current power functional theory (black solid
  line). Data taken at time $t=0.1\tau$ after switching off the
  external potential. The inset in the top panel shows the time evolution
  of $K$ (prefactor in $P_{\rm{exc}}$) as a function of the scale time $t/\tau$.
  In panels b) we show the same data as in panels
  a) for a system with $N=20$ and $H=30$.}
\label{fig1}
\end{figure}

Fig.~\ref{fig1} shows the density and the excess adiabatic and superadiabatic
force density profiles of systems with $N=15$ (a) and $N=20$ (b) at time
$t=0.1\tau$, with $\tau=L^2\gamma/(k_{\rm B}T)$, and $H/L=30)$. 
The excess adiabatic and superadiabatic force densities are of the same
order of magnitude. In (a) superadiabatic and adiabatic forces are out
of phase, whereas the opposite is true in (b). These examples highlight
the important contribution of $P_t^{\rm exc}$ to the force balance: The
magnitude of the superadiabatic force is not negligible and its structure
is nontrivial. The agreement between PFT and BD is excellent in all cases
analysed.

The insets of Fig.~\ref{fig1} show the prefactor $K$ of $P_t^{\rm exc}$, which
measures the magnitude of $I_{\rm sup}$, as a function of time for
systems with $N=15$ and $20$ ($H/L=30$ in both cases).  As expected,
the superadiabatic force vanishes for $t=0$ (since the system is at
equilibrium at $t=0$) and reaches a plateau as time evolves due to the
saturation of memory effects.

The reformulation of PFT in terms of the gradient of the velocity
field, as presented here, is amenable to the study of stress-stress
and strain rate-strain rate correlation functions via functional
differentiation, and corresponding nonequilibrium Ornstein-Zernike
relations \cite{brader13noz1,brader13noz2}.

In future work, the explicit study of memory effects is an important
topic. Higher (than bilinear) order contributions to $P_t^{\rm exc}$
can be systematically constructed from combinations of the scalars
$\nabla\cdot\vel$ and $(\nabla\times\vel)^2$. The resulting
nonequilibrium forces go beyond the viscous forces that follow from
\eqref{EQPtexc_local}. Work along these lines will be presented
elsewhere \cite{tobinico}. 
Further possible interesting applications are the application to
gravitational collapse \cite{bleibel2016} of monolayers and active
microrheology \cite{gruber2016fuchs}.

\acknowledgments We thank N. Stuhlm\"uller, T. Eckert, and
L.~Treffenst\"adt for useful discussions. This work is supported by
the German Research Foundation (DFG) via SCHM 2632/1-1.

\end{document}